\documentstyle[psfig,aps,prc,eqsecnum]{revtex}

\begin{document}
\draft
\title{Peripheral N$\alpha $ scattering: a tool for identifying the two pion
exchange component of the NN potential}
\author{L. A. Barreiro, R. Higa, C. L. Lima, and M. R. Robilotta}
\address{Nuclear Theory and Elementary Particle Phenomenology Group\\
Instituto de F\'{\i}sica, Universidade de S\~{a}o Paulo,\\
Caixa Postal 66318, 05315-970, S\~{a}o Paulo, SP, Brazil}
\date{\today
}
\maketitle

\begin{abstract}
We study elastic N$\alpha $ scattering and produce a quantitative
correlation between the range of the effective potential and the energy of
the system. This allows the identification of the waves and energies for
which the scattering may be said to be peripheral. We then show that the
corresponding phase shifts are sensitive to the tail of the NN potential,
which is due to the exchange of two pions. However, the present
uncertainties in the experimental phase shifts prevent the use of N$\alpha $
scattering to discriminate the existing models for the NN interaction.
\end{abstract}


\section{Introduction}

In the present understanding of the $NN$ interaction, long-range effects are
ascribed to single pion exchanges, intermediate components are associated
with exchanges of two and three pions, $\rho $'s and $\omega $, as well as
baryon excitations, such as the $\Delta $. At short distances, quark
dynamics is supposed to dominate.

The long-range one-pion exchange potential (OPEP) became well established in
the 1960's \cite{hamada} as a necessary component of all realistic forces.
This achievement was followed by an effort to determine the next layer of
the interaction, associated with the two-pion exchange potential (TPEP), and
various strategies were proposed to describe it. Early in 1971, Brown and
Durso \cite{brown-durso} pointed out that this component of the force is
directly related to off-shell $\pi N$ scattering and hence strongly
influenced by chiral symmetry. In a subsequent paper, Chemtob, Durso, and
Riska \cite{chemtob} explored this relationship and investigated the
phenomenological features of the TPEP obtained by means of dispersion
relations. A very important step in this research program was the
construction of the Paris potential \cite{paris}, where dispersion relations
were used to relate empirical information about the $\pi N$ process to the
intermediate part of the force. This potential is rather successful in
describing experimental data.

The intermediate part of the potential may also be studied in the framework
of field theory. In this case, one writes down a Lagrangian involving the
relevant degrees of freedom and then evaluate a certain number of Feynman
diagrams. This leads to an amplitude which is afterwards transformed into a
potential. An important early work along this line was that of Partovi and
Lomon \cite{partovi}, who considered a Lagrangian containing just pions and
nucleons with a pseudoscalar (PS) coupling and evaluated the box and crossed
box diagrams contributing to the exchange of two uncorrelated pions. A
detailed study of the same diagrams using a pseudovector (PV) coupling was
performed later by Zuilhof and Tjon \cite{zuilhof}. The inclusion of the
exchanges of resonances and baryon excitations led to the construction of
the Bonn potential \cite{machleidt1,machleidt2}, which is also quite
successful in reproducing experimental data.

In a parallel line of development, several phenomenological potentials were
constructed, which are also able to account for experimental information
with the aid of adjusted parameters \cite{tourreil1,tourreil2,pandha,wiringa}%
.

An interesting feature of all approaches to the intermediate part of the
potential is that their theoretical reliability decreases as one moves from
the outer to the inner region. The intermediate range interactions exhibit a
marked spatial hierarchy. The spatial features of a given interaction are
determined by the mass exchanged in the $t$ channel. In the intermediate
part of the potential, the lightest system that can be exchanged involves
just two pions and has a mass of about 300 MeV. Other important effects,
associated with resonances such as the $\rho $ and the $\omega $ are short
ranged, because these states have masses around 750 MeV. When field theory
is used, predictions for the inner parts of the mesonic sector of the
potential are heavily influenced by form factors, bringing a lot of
uncertainty to calculations. In the case of dispersion relations, on the
other hand, predictions for the inner region are based on data which need to
be extrapolated far away from their experimental region.

Recently there has been a renewal in the interest on the intermediate part
of the $NN$ interaction, motivated by the realization that chiral symmetry
provides a suitable theoretical framework for the calculation of a strong
process \cite{bira1,celenza,friar,carocha}. In its minimal version, the
chiral intermediate potential is based on a system containing just pions and
nucleons. The theoretical foundations of this part of the potential are
rather well established and it is reasonable to expect that it should become
a standard ingredient of any modern $NN$ potential. A shortcoming of the
minimal chiral potential is that it fails to reproduce experimental
information in the case of the intermediate $\pi N$ amplitude entering in
the TPEP. In order to overcome this difficulty, one may extend the chiral
model, so as to encompass other degrees of freedom. This possibility was
recently considered by Ord\'{o}\~{n}ez, Ray, and van Kolck \cite{bira2}, who
have shown that the inclusion of $\Delta $'s in the model improves its
predictive power. Alternatively, one may choose to introduce the empirical
information that is missing in the intermediate $\pi N$ amplitude in a model
independent way, with the help of the H\"{o}hler-Jacob-Strauss \cite{h"hler}
subthreshold coefficients \cite{mane1,mane2}. This led to a TPEP which
yields a satisfactory description of $NN$ data for waves and energies
associated with peripheral scattering \cite{mane3,kaiser}.

In spite of the considerable amount of activity related to the intermediate
part of the $NN$ potential in the last twenty-five years, no consensus was
reached about the fine details of this component of the force. The dynamical
content of the various models is not uniform and the profile functions they
yield for different components of the potential do not agree. This picture
poses the problem of defining criteria for establishing the merits and
shortcomings of the various existing potentials, so that a choice can be
produced. An obvious criterion of choice is the ability a given potential
has of reproducing experimental data. This possibility has been on for a
long time and does not work as a distinguishing criterion, because all
existing modern potentials are able to explain well experimental $NN$ phase
shifts, but do so with the help of several free parameters that are adjusted 
{\it ad hoc}. Another problem about relying on $NN$ observables is that the
OPEP contributes to all channels and waves, making it difficult to isolate
unambiguously the contribution of intermediate range dynamics. It would
therefore be interesting to find ways of obtaining information about the
intermediate part of the potential directly from empirical data.

In this work we speculate about the possibility of obtaining such an
information from the study of nucleon-alpha $(N\alpha )$ scattering. The $%
\alpha $ particle is a rather suitable system for the study of the
intermediate part of the nuclear interaction, because it is a boson that has
no spin and isospin and hence cannot couple to a single pion. This means
that the outer part of the nuclear potential surrounding the $\alpha $
involves two uncorrelated pions. The scalar-isoscalar channel of the
two-pion exchange $NN$ potential has strong central and spin-orbit
components, and the same happens with the $N\alpha $ effective interaction.
This picture has empirical support, since $N\alpha $ experimental phase
shifts show that central and spin-orbit effects are very important \cite
{exper1,exper2,exper3,exper4}. Therefore we may expect that the peripheral
scattering of nucleons by the $\alpha $ should be heavily dominated by
two-pion exchange process and hence reflect the various approaches adopted
in different $NN$ models. When low-energy protons are used as probes, the
Coulomb barrier prevents short distance interactions. Thus the effects due
to the nuclear force show themselves as deviations from Coulomb amplitudes,
similarly to what happens in low-energy nucleus-nucleus collisions \cite
{pb-pb,dirceu}. For instance, in sub-Coulomb Pb+Pb Mott scattering the
intermediate range $NN$ interaction accounts for deviations observed
experimentally \cite{pb-pb}. In the case of low-energy neutrons, on the
other hand, angular momentum may be used to select the various regions of
the potential.

This work is organized as follows. The effective $N\alpha $ potential is
defined in sect. \ref{sec2} and explicitly constructed in sect. \ref{sec3}.
In sect. \ref{sec4} we display our dynamical equations, which are based on
the variable phase method. Finally, in sect. \ref{sec5} we discuss our
results.

\section{Dynamics}

\label{sec2}

In this section we present briefly the main equations used in this work. In
a complete treatment of the $N\alpha $ scattering problem, one has to deal
with a rather complex five-body system. However, here we are interested in
estimating the effects of the tail of the $NN$ interaction, which manifests
themselves at large distances and high values of the angular momentum. We
therefore assume that the $\alpha $ subsystem remains undisturbed during the
interaction.

The wave function for the four-body nuclear system $^4$He can be written, in
terms of the spatial, spin, and isospin degrees of freedom, as 
\begin{eqnarray}
|N_1\cdot \cdot \cdot N_4\rangle &=&|{\bf r}_1\cdot \cdot \cdot {\bf r}%
_4\rangle \otimes |\text{spin}\rangle \otimes |\text{isospin}\rangle 
\nonumber \\
\ &=&|{\bf R}\rangle |\alpha \rangle ,  \label{II.A1}
\end{eqnarray}
where the collective variable ${\bf R}$ is given in terms of the individual
coordinates ${\bf r}_i$: 
\begin{equation}
{\bf R}=\frac 14\sum_{i=1}^4{\bf r}_i  \label{II.A2}
\end{equation}
and $|\alpha \rangle $ represents the $^4$He ground state. This wave
function is known to have $T=0,$ $J^\pi =0^{+}$, and both $S$ and $D$
spatial components \cite{nipon1}. In order to simplify the calculation we
neglect $D$ waves, adopt a Gaussian structure for the spatial wave function 
\cite{nipon1} and write 
\begin{equation}
|\alpha \rangle =N_\alpha \exp \left( -\frac \alpha 2\sum_{i>j=1}^4{\bf r}%
_{ij}^{\text{ }2}\right) |\chi _\alpha \rangle ,  \label{II.A3}
\end{equation}
where ${\bf r}_{ij}=($ ${\bf r}_i-{\bf r}_j)$, $N_\alpha $ is a
normalization constant, $\alpha $ is a parameter extracted from Ref. \cite
{nipon1}, and $|\chi _\alpha \rangle $ is the spin-isospin wave function.

The dynamics of the four body system is determined by the Hamiltonian 
\begin{eqnarray}
H_4 &=&\sum_{i=1}^4-\frac{\nabla _i^2}{2m}+\sum_{i>j=1}^4\Theta _{ij} 
\nonumber \\
&=&-\frac{\nabla _{\text{R}}^2}{2M_\alpha }+H_\alpha ,  \label{II.A11}
\end{eqnarray}
where $\Theta _{ij}=k{\bf r}_{ij}^{\text{ }2}/2$ is the harmonic potential
and $k=4\alpha ^2/m$.

In order to isolate the motion of the center of mass, it is useful to use
the Jacobi variables, given generically by 
\begin{equation}
{\mbox{\boldmath $\rho$}_i}=\sum_{j=1}^4a_{ij}{\bf r}_j,
\end{equation}
where the coefficients $a_{ij}$ are summarized in the matrix 
\begin{equation}
{\mbox{\boldmath $a$}}=\left( 
\begin{array}{cccc}
\sqrt{1/2} & -\sqrt{1/2} & 0 & 0 \\ 
\sqrt{1/6} & \sqrt{1/6} & -\sqrt{2/3} & 0 \\ 
\sqrt{1/12} & \sqrt{1/12} & \sqrt{1/12} & -\sqrt{3/4}
\end{array}
\right) .
\end{equation}

In terms of these new coordinates, the Hamiltonian for the isolated
four-body system is

\begin{equation}
H_4=-\frac{\nabla _{\text{R}}^2}{2M_\alpha }+H_\alpha ,  \label{II.A12a}
\end{equation}
where $-\nabla _{\text{R}}^2/2M_\alpha $ describes the center-of-mass motion
and 
\begin{equation}
H_\alpha =\sum_{i=1}^3\left[ -\frac{\nabla _{_{\rho _i}}^2}{2m_i}+2k{%
\mbox{\boldmath $\rho$}}_i^2\right]  \label{II.A13}
\end{equation}
is the intrinsic four-body Hamiltonian. The Schr\"{o}dinger equation for the 
$\alpha $ particle is 
\begin{equation}
H_\alpha \mid \alpha \rangle =E_\alpha \mid \alpha \rangle ,  \label{II.A15}
\end{equation}
where $E_\alpha =M_\alpha -4m$ and $|\alpha \rangle $ is now given by 
\begin{equation}
\left| \alpha \right\rangle =\left( \frac{4\alpha }\pi \right) ^{9/4}\exp
\left( -2\alpha \sum_{i=1}^3{\mbox{\boldmath $\rho$}}_i^{\text{ }2}\right)
|\chi _\alpha \rangle .
\end{equation}

The strong interaction between the incoming particle, hereafter labeled by $%
o,$ and the nucleon $i$ within the $\alpha $ particle is denoted by $V_{oi}(%
{\bf r}_{oi})$. The study of the effects of this interaction is the main
object of this work.

The Schr\"{o}dinger equation for this five nucleon system is

\begin{equation}
H_5|N_o\cdot \cdot \cdot N_4\rangle =E|N_o\cdot \cdot \cdot N_4\rangle ,
\label{II.B3a}
\end{equation}
where 
\begin{eqnarray}
H_5 &=&\sum_{i=0}^4-\frac{\nabla _i^2}{2m}+\sum_{i>j=1}^4\Theta
_{ij}+\sum_{i=1}^4V_{oi}  \nonumber \\
&=&H_\alpha -\frac{\nabla _{\text{R}}^2}{2M_\alpha }-\frac{\nabla _o^2}{2m}%
+\sum_{i=1}^4V_{oi}.  \label{II.B5a}
\end{eqnarray}

Approximating the five-body wave function by a product of the different
clusters constituting the system 
\begin{equation}
|N_o\cdot \cdot \cdot N_4\rangle \simeq |N_o,{\bf R}\rangle |\alpha \rangle ,
\label{II.B5}
\end{equation}
we can rewrite the dynamical equations as 
\begin{eqnarray*}
H_5|N_o\cdot \cdot \cdot N_4\rangle &\simeq &\left[ E_\alpha -\frac{\nabla _{%
\text{R}}^2}{2M_\alpha }-\frac{\nabla _o^2}{2m}+\sum_{i=1}^4V_{oi}\right]
|N_o,{\bf R}\rangle |\alpha \rangle \\
&=&E|N_o,{\bf R}\rangle |\alpha \rangle .
\end{eqnarray*}

Introducing two new Jacobi variables

\begin{eqnarray}
{\bf s} &=&\frac 15({\bf r}_o+4{\bf R}),  \nonumber \\
{\bf x} &=&{\bf r}_o-{\bf R},  \label{II.B9}
\end{eqnarray}
we have 
\begin{equation}
\left[ -%
{\displaystyle {\nabla _{\text{s}}^2 \over 2(M_\alpha +m)}}
-%
{\displaystyle {\nabla _{\text{x}}^2 \over 2\mu }}
+\sum_{i=1}^4V_{oi}\right] |N_o,{\bf R}\rangle |\alpha \rangle =(E-E_\alpha
)|N_o,{\bf R}\rangle |\alpha \rangle ,  \label{II.B11}
\end{equation}
where 
\begin{equation}
\mu =\frac{M_\alpha m}{M_\alpha +m}\simeq \frac 45m.  \label{II.B13}
\end{equation}

Writing down explicitly the projectile wave function as 
\begin{equation}
|N_o\rangle =|{\bf r}_o\rangle |\chi _o\rangle
\end{equation}
and going to the C.M. frame of the five-body system, we obtain

\begin{equation}
\left[ -\frac{\nabla _{\text{x}}^2}{2\mu }+\sum_{i=1}^4V_{oi}\right] |{\bf x}%
\rangle |\chi _o\rangle |\alpha \rangle =E_{\text{x}}|{\bf x}\rangle |\chi
_o\rangle |\alpha \rangle ,  \label{II.B27}
\end{equation}
where $E_{\text{x}}=E-E_\alpha $. Multiplying this equation by $\langle
\alpha |$ and integrating over the $\alpha $ coordinates, we get 
\begin{equation}
\left[ -\frac{\nabla _{\text{x}}^2}{2\mu }+W({\bf x})\right] |{\bf x}\rangle
|\chi _o\rangle =E_x|{\bf x}\rangle |\chi _o\rangle ,
\end{equation}
where 
\begin{equation}
W({\bf x})=\langle \alpha |\sum_{i=1}^4V_{oi}({\bf r}_{oi})\mid \alpha
\rangle  \label{II.B29}
\end{equation}
is the effective $N-$ $\alpha $ potential.

\section{The effective potential}

\label{sec3}

The two-body strong potential $V_{oi}$ is written as 
\begin{equation}
V_{oi}=\sum_{S,T}V_{oi}^{TS}({\bf r}_{oi})P_{oi}^TP_{oi}^S,  \label{II.C1}
\end{equation}
where the indices $T$ and $S$ represent the total isospin and spin of the $%
NN $ system whereas $P_{oi}^T$ and $P_{oi}^S$ are the corresponding
projection operators. The explicit forms of the isospin operators are 
\begin{equation}
P_{oi}^0=\frac 14\left[ 1-{\mbox{\boldmath $\tau$}}^{(o)}\cdot {%
\mbox{\boldmath $\tau$}}^{(i)}\right] ,  \label{II.C3}
\end{equation}
\begin{equation}
P_{oi}^1=\frac 14\left[ 3+{\mbox{\boldmath $\tau$}}^{(o)}\cdot {%
\mbox{\boldmath $\tau$}}^{(i)}\right] ,  \label{II.C5}
\end{equation}
and similar expressions hold for the spin degrees of freedom.

The various radial components $V_{oi}^{TS}$ have the following structure 
\begin{eqnarray}
V_{oi}^{TS} &=&V_C^{TS}({\bf r}_{oi})+V_{LS}^{TS}({\bf r}_{oi}){\bf L}%
_{oi}\cdot \left( \frac{{\mbox{\boldmath $\sigma$}}^{(o)}}2+\frac{{%
\mbox{\boldmath $\sigma$}}^{(i)}}2\right)  \nonumber \\
&&+V_T^{TS}({\bf r}_{oi})\left( 3{\mbox{\boldmath $\sigma$}}^{(o)}\cdot {\bf 
\hat{r}}_{oi}{\mbox{\boldmath $\sigma$}}^{(i)}\cdot {\bf \hat{r}}_{oi}-{%
\mbox{\boldmath $\sigma$}}^{(o)}\cdot {\mbox{\boldmath $\sigma$}}%
^{(i)}\right)  \nonumber \\
&&+\text{small components.}
\end{eqnarray}
In this expression, ${\bf L}_{oi}$ is the orbital angular momentum for the
pair $oi$, whereas $V_C^{TS}$, $V_{LS}^{TS},$ and $V_T^{TS}$ represent the
central, spin-orbit, and tensor components of the potential, respectively.

In the evaluation of the effective potential, one needs the expectation
value of $V_{oi}$ between the wave functions of the $\alpha $ particle.
Noting that $\langle \chi _\alpha |{\mbox{\boldmath $\sigma$}}^{(i)}|\chi
_\alpha \rangle =\langle \chi _\alpha |{\mbox{\boldmath $\tau$}}^{(i)}|\chi
_\alpha \rangle =0$, we have 
\begin{equation}
\langle \chi _\alpha |V_{oi}({\bf r}_{oi})|\chi _\alpha \rangle =V_C({\bf r}%
_{oi})+V_{LS}({\bf r}_{oi}){\bf L}_{oi}\cdot \frac{{\mbox{\boldmath $\sigma$}%
}^{\left( o\right) }}2,  \label{II.C9}
\end{equation}
where 
\begin{equation}
V_C({\bf r}_{oi})=\frac 1{16}\left[ V_C^{00}({\bf r}_{oi})+3V_C^{10}({\bf r}%
_{oi})+3V_C^{01}({\bf r}_{oi})+9V_C^{11}({\bf r}_{oi})\right]  \label{II.C10}
\end{equation}
and 
\begin{equation}
V_{LS}({\bf r}_{oi})=\frac 14\left[ V_{LS}^{01}({\bf r}_{oi})+3V_{LS}^{11}(%
{\bf r}_{oi})\right] .  \label{II.C10a}
\end{equation}

The strong effective potential is then given by 
\begin{eqnarray}
W({\bf x}) &=&\sum_{i=1}^4\left( \frac{4\alpha }\pi \right) ^{9/2}\int d{%
\mbox{\boldmath $\rho$}}_1d{\mbox{\boldmath $\rho$}}_2d{%
\mbox{\boldmath
$\rho$}}_3\exp \left( -2\alpha \sum_{j=1}^3{\mbox{\boldmath $\rho$}}_j^{%
\text{ }2}\right)  \nonumber \\
&&\ \times \left[ V_C({\bf r}_{oi})+V_{LS}({\bf r}_{oi}){\bf L}_{oi}\cdot 
\frac{{\mbox{\boldmath
$\sigma$}}^{\left( o\right) }}2\right] \exp \left( -2\alpha \sum_{j=1}^3{%
\mbox{\boldmath $\rho$}}_j^{\text{ }2}\right) .  \label{II.C12}
\end{eqnarray}

In the evaluation of this expression, it is convenient to note that 
\begin{equation}
{\bf r}_i={\bf R}+\sum_{j=1}^3a_{ij}^T{\mbox{\boldmath $\rho$}}_j
\end{equation}
and hence 
\begin{equation}
{\bf r}_{oi}={\bf x}-\sum_{j=1}^3a_{ij}^T{\mbox{\boldmath $\rho$}}_j.
\end{equation}

The angular momentum operator ${\bf L}_{oi}$ is given by ${\bf L}_{oi}={\bf r%
}_{oi}\times {\bf p}_{oi}$, where ${\bf p}_{oi}=-i\nabla _{oi}$. The use of
the chain rule allows us to write 
\begin{eqnarray}
{\bf L}_{oi} &=&\left( {\bf x}-\sum_{j=1}^3a_{ij}^T{\mbox{\boldmath $\rho$}}%
_j\right) \times \left( \frac 14{\bf p}_{\text{x}}-\sum_{k=1}^3a_{ik}^T{\bf p%
}_k\right)  \nonumber \\
\ &=&\frac 14{\bf x}\times {\bf p}_{\text{x}}-\frac 14\sum_{j=1}^3a_{ij}^T({%
\mbox{\boldmath $\rho$}}_j\times {\bf p}_{\text{x}})-\sum_{k=1}^3a_{ik}^T(%
{\bf x}\times {\bf p}_k)  \nonumber \\
&&+\sum_{k=1}^3\sum_{j=1}^3a_{ij}^Ta_{ik}^T({\mbox{\boldmath $\rho$}}%
_j\times {\bf p}_k).
\end{eqnarray}

When the operator ${\bf p}_k$ acts on the spatial wave function, we have 
\begin{equation}
{\bf p}_k\exp \left( -2\alpha \sum_{i=1}^3{\mbox{\boldmath $\rho$}}%
_i^2\right) =-i(-4\alpha ){\mbox{\boldmath $\rho$}}_k\exp \left( -2\alpha
\sum_{i=1}^3{\mbox{\boldmath $\rho$}}_i^{\text{ }2}\right) .  \label{II.C21}
\end{equation}

Defining ${\bf L}_{\text{x}}={\bf x}\times {\bf p}_{\text{x}},$ we write 
\begin{equation}
W({\bf x})=W_C({\bf x})+\frac 14W_{LS}({\bf x}){\bf L}_{\text{x}}\cdot \frac{%
{\mbox{\boldmath $\sigma$}}^{(o)}}2.  \label{II.C22}
\end{equation}

The Gaussian structure of the wave function allows several integrations to
be performed analytically and we obtain

\begin{equation}
W_C({\bf x})=4\left( \frac{16\alpha }{3\pi }\right) ^{1/2}\frac 1x%
\int_0^\infty duuV_C(u)\left\{ \exp \left[ -\frac{16\alpha }3(x-u)^2\right]
-\exp \left[ -\frac{16\alpha }3(x+u)^2\right] \right\}  \label{II.C25}
\end{equation}
and 
\begin{eqnarray}
W_{LS}({\bf x}) &=&4\left( \frac{16\alpha }{3\pi }\right) ^{1/2}\frac 1{x^2}%
\int_0^\infty duuV_{LS}(u)\times  \label{II.C25a} \\
&&\left\{ \left( u-\frac 3{32\alpha x}\right) \exp \left[ -\frac{16\alpha }3%
(x-u)^2\right] +\left( u+\frac 3{32\alpha x}\right) \exp \left[ -\frac{%
16\alpha }3(x+u)^2\right] \right\} .  \nonumber
\end{eqnarray}

For a channel with total angular momentum ${\bf J}$ and orbital angular
momentum ${\bf L},$ we obtain the following form for the effective
potential: 
\begin{equation}
W_{JL}(x)=W_C(x)+\frac 18\left[ J(J+1)-L(L+1)-\frac 34\right] W_{LS}(x).
\label{II.C26}
\end{equation}
Thus we have 
\begin{eqnarray}
W_{L-\frac 12L}(x) &=&W_C(x)-\frac 18(L+1)W_{LS}(x),  \label{l1} \\
W_{L+\frac 12L}(x) &=&W_C(x)+\frac 18LW_{LS}(x).  \label{l2}
\end{eqnarray}

\section{The variable phase approach}

\label{sec4}

We are interested in the properties of the tail of the $NN$ interaction and
hence it is convenient to calculate the $N\alpha $ phase shifts by means of
the so-called variable phase method, as described by Calogero \cite{calogero}%
. This method is fully equivalent to the Schr\"{o}dinger formalism, and has
the advantage of providing a clear picture of the spatial influence of the
potential over the phase shift and of its relationship with the centrifugal
barrier. For the reader's convenience, we hereafter collect the main
formulas needed in this calculation. As we disregard tensor interactions,
there are only uncoupled channels. The wave $u_{JL}(r)$, representing a
system with total and orbital angular momenta $J$ and $L$, respectively, is
written in terms of the radial Green's functions as 
\begin{equation}
u_{JL}(r)=c_{JL}(k,r)\hat{j}_L(kr)-s_{JL}(k,r)\hat{n}_L(kr),  \label{II.D1}
\end{equation}
where $k=\sqrt{2\mu E_x}$, $\hat{j}_L(kr)$ and $\hat{n}_L(kr)$ are the usual
spherical Bessel and Neumann functions multiplied by $kr$ whereas $%
c_{JL}(k,r)$ and $s_{JL}(k,r)$ are defined in terms of $V_{JL}(r)$ as
follows: 
\begin{equation}
c_{JL}(k,r)=1-\frac 1k\int_0^rdr^{\prime }V_{JL}(r^{\prime })\hat{j}%
_L(kr^{\prime })u_{JL}(r^{\prime }),  \label{II.D2}
\end{equation}
\begin{equation}
s_{JL}(k,r)=-\frac 1k\int_0^rdr^{\prime }V_{JL}(r^{\prime })\hat{n}%
_L(kr^{\prime })u_{JL}(r^{\prime }).  \label{II.D3}
\end{equation}
The phase shift $\delta _{JL}\left( k\right) $ is related to these functions
by 
\begin{equation}
\tan \delta _{JL}\left( k\right) =\lim_{r\rightarrow \infty }\frac{%
s_{JL}(k,r)}{c_{JL}(k,r)}.  \label{II.D4}
\end{equation}

This motivates the definition of a variable phase $\delta _{JL}\left(
k,r\right) $ through 
\begin{equation}
\delta _{JL}\left( k,r\right) =\arctan \left[ \frac{s_{JL}(k,r)}{c_{JL}(k,r)}%
\right] .  \label{II.D5}
\end{equation}
This function vanishes at the origin \cite{calogero} and becomes the phase
shift when $r$ approaches infinity. It is determined dynamically by means of
a first-order differential equation, obtained by differentiating Eq. (\ref
{II.D5}), using Eqs. (\ref{II.D2}), (\ref{II.D3}), and manipulating the
result to obtain 
\begin{equation}
\frac d{dr}\delta _{JL}\left( k,r\right) =-\frac 1kW_{JL}(r)P_L^2\left[
kr,\delta _{JL}\left( k,r\right) \right] ,  \label{II.D6}
\end{equation}
where $P_L$ is the uncoupled structure function, given by 
\begin{equation}
P_L\left[ kr,\delta _{JL}\left( k,r\right) \right] =\cos \left[ \delta
_{JL}\left( k,r\right) \right] \hat{j}_L(kr)-\sin \left[ \delta _{JL}\left(
k,r\right) \right] \hat{n}_L(kr).  \label{II.D7}
\end{equation}
The integral expression for the variable phase is 
\begin{equation}
\delta _{JL}\left( k,r\right) =-\frac 1k\int_0^rdr^{\prime }V_{JL}(r^{\prime
})P_L^2\left[ kr^{\prime },\delta _{JL}\left( k,r^{\prime }\right) \right] .
\label{II.D8}
\end{equation}

For future use, we define the ratio $\rho _{JL}(k,r)$ as 
\begin{equation}
\rho _{JL}(k,r)=\frac{\delta _{JL}\left( k,r\right) }{\delta _{JL}\left(
k,\infty \right) }.  \label{e48}
\end{equation}
It represents the fraction of the phase shift that is generated by the part
of the potential between the origin and the point $r$.

\section{Results and discussion}

\label{sec5}

The purpose of this work is to discuss the possibility of using $N\alpha $
scattering to study the tail of the two-pion exchange nucleon-nucleon
potential. As stressed previously, the $\alpha $ particle is a spin-isospin
scalar and hence it is very suited to the study of the intermediate-range
part of the nuclear interaction. We therefore consider several realistic $NN$
potentials \cite{paris,machleidt1,tourreil1,tourreil2,wiringa,mane2} and
fold them into the $\alpha $ wave function, in order to obtain effective
interactions, which are then used in the Schr\"{o}dinger equation. The
effective $N\alpha $ potential is dominated by scalar and isoscalar
exchanges, giving rise to central and spin-orbit contributions, displayed in
Fig. \ref{fig1}. In all cases, as expected, the spin-orbit component falls
faster than the central one at large distances. Both components are negative
there and hence, according to Eqs. (\ref{l1}) and (\ref{l2}), they interfere
destructively when $J=L-\frac 12$ and constructively when $J=L+\frac 12$.
This means that phase shifts for peripheral waves of $L$ are positive and
larger for the latter class of waves.

We are interested in effects of the tail of the potential and hence it is
important to establish, for each wave, a correlation between the energy and
the region of the potential that determines the phase shifts. With that goal
in mind, we have employed the variable phase method to evaluate the ratio $%
\rho _{JL}(k,r)$, defined in Eq. (\ref{e48}), for various energies, in order
to determine the radii $R_5$ and $R_{10}$ for which $\rho _{JL}(k,r)$ is
always less than 5 and 10 \%, respectively. This means that 95\% of the
phase shift is generated in the region where $r>R_5$ and, similarly, 90\%
for $r>R_{10}$. Results for the Argonne potential \cite{wiringa} are
presented in Table \ref{tab1}, where it is possible to notice some
interesting features. The first of them is that, for a given value of $L$,
the phases are slightly more sensitive to the short-range part of the
potential for the channel with $J=L-\frac 12$. This feature of the problem
can be fully appreciated in Fig. \ref{fig2}, that shows the ratios $\rho
_{F_{5/2}}$ and $\rho _{F_{7/2}}$ at 15 MeV, which are representative of the
other peripheral waves. The behavior of these curves are determined by the
centrifugal barrier, responsible for the plateau close to the origin,
combined with an effective potential which is repulsive at short distances
and then attractive.

The $\alpha $ particle has a rms of about 1.6 fm and, in a conservative
approach, we may define the tail of the effective potential as beginning at
2.0 fm. Table \ref{tab1} then informs us that such a tail of the $N\alpha $
potential determines more than 95\% of the phase shifts $\delta _{G_{7/2}}$, 
$\delta _{G_{9/2}}$ up to 50 MeV and $\delta _{H_{9/2}},$ $\delta
_{H_{11/2}} $ up to more than 80 MeV.

In Fig. \ref{fig3} we display the phase shifts for waves with $2\leq L\leq 5$%
. Inspecting the figures corresponding to the waves $F$, $G$, and $H$, one
learns that the various potentials provide a coherent pattern, since all
curves have the same general trends. The phases are sensitive to the tail of
the effective potential and hence one could, in principle, discriminate the
medium range features of the $NN$ interactions. In Fig. \ref{fig3} we also
include the available experimental values, taken from Refs. \cite
{exper1,exper2,exper3,exper4}. The picture provided by these data, claimed
to be precise, does not allow conclusions, for there are big discrepancies
among phase shift analyses made by the various authors. So a drastic
improvement in the precision of $N\alpha $ phase shifts is needed before
they could be used to discriminate $NN$ potentials.

As a final comment, we would like to stress that our calculation is a very
simple one and a more detailed study is required on the theoretical side.
One of the points that deserves further attention concerns the construction
of the $\alpha $ wave function, that should include $D$ components and, if
possible, be calculated using realistic $NN$ potentials. Although it is
reasonable to assume that intermediate states involving the $^3$Li$_5$, $^2$%
He$_5$, or $\alpha $ excitations contribute little to peripheral scattering,
it is also important to clarify this aspect of the problem. Finally, it
would also be important to establish the quantitative role of the two-pion
exchange three-body force, since its range is comparable to that of the
TPEP. We are now investigating some of these questions and prefer to take
the results of the present work as indicating reliably only that $N\alpha $
scattering allows the discrimination of the medium-range content of $NN$
potentials. In order to be able to select one or a class of potentials as
being better than others one needs both more precise calculations and
experimental phase shifts.

\begin{center}
{\bf ACKNOWLEDGMENT}
\end{center}

L. A. B. and R. H. were supported by Funda\c {c}\~{a}o de Amparo \`{a}
Pesquisa do Estado de S\~{a}o Paulo (FAPESP).

\newpage

%
%
\begin{table}[tbp]
\centering
\begin{tabular}{ccccccccccccc}
& \multicolumn{2}{c}{$F_{5/2}$} & \multicolumn{2}{c}{$F_{7/2}$} & 
\multicolumn{2}{c}{$G_{7/2}$} & \multicolumn{2}{c}{$G_{9/2}$} & 
\multicolumn{2}{c}{$H_{9/2}$} & \multicolumn{2}{c}{$H_{11/2}$} \\ 
E & $R_5$ & $R_{10}$ & $R_5$ & $R_{10}$ & $R_5$ & $R_{10}$ & $R_5$ & $R_{10}$
& $R_5$ & $R_{10}$ & $R_5$ & $R_{10}$ \\ \hline
5 & 2.12 & 3.26 & 2.96 & 3.10 & 3.24 & 3.48 & 3.10 & 3.34 & 3.62 & 3.96 & 
3.50 & 3.86 \\ 
&  &  &  &  &  &  &  &  &  &  &  &  \\ 
10 & 1.96 & 3.22 & 2.24 & 3.06 & 3.18 & 3.38 & 3.04 & 3.24 & 3.48 & 3.78 & 
3.36 & 3.68 \\ 
&  &  &  &  &  &  &  &  &  &  &  &  \\ 
15 & 1.84 & 2.22 & 2.00 & 3.04 & 3.14 & 3.32 & 2.98 & 3.18 & 3.38 & 3.66 & 
3.26 & 3.54 \\ 
&  &  &  &  &  &  &  &  &  &  &  &  \\ 
20 & 1.74 & 2.02 & 1.86 & 3.02 & 3.10 & 3.26 & 2.94 & 3.12 & 3.30 & 3.56 & 
3.18 & 3.44 \\ 
&  &  &  &  &  &  &  &  &  &  &  &  \\ 
30 & 1.56 & 1.78 & 1.68 & 1.98 & 3.06 & 3.20 & 2.88 & 3.02 & 3.20 & 3.40 & 
3.06 & 3.28 \\ 
&  &  &  &  &  &  &  &  &  &  &  &  \\ 
40 & 1.40 & 1.58 & 1.52 & 1.74 & 2.14 & 3.14 & 2.82 & 2.96 & 3.12 & 3.30 & 
2.96 & 3.16 \\ 
&  &  &  &  &  &  &  &  &  &  &  &  \\ 
50 & 1.26 & 1.40 & 2.23 & 2.95 & 1.94 & 2.34 & 2.80 & 2.92 & 3.06 & 3.22 & 
2.90 & 3.08 \\ 
&  &  &  &  &  &  &  &  &  &  &  &  \\ 
60 & 1.08 & 1.20 & 1.97 & 2.91 & 1.80 & 2.06 & 2.00 & 2.88 & 3.02 & 3.16 & 
2.84 & 3.00 \\ 
&  &  &  &  &  &  &  &  &  &  &  &  \\ 
70 & 0.78 & 0.86 & 1.12 & 1.26 & 1.68 & 1.88 & 1.84 & 2.86 & 3.00 & 3.12 & 
2.80 & 2.94 \\ 
&  &  &  &  &  &  &  &  &  &  &  &  \\ 
80 & 0.86 & 1.06 & 0.98 & 1.08 & 1.56 & 1.74 & 1.72 & 1.98 & 2.26 & 3.08 & 
2.76 & 2.90
\end{tabular}
\caption{Radii $R_5$ and $R_{10}$, in fm, as function of the energy E, in
MeV, for which the normalized phase shift, calculated for the Argonne
potential, is always less than 5\% and 10\%, respectively. It is worth
noting the non monotonic behavior of the $F_{5/2}$ wave, which is due to a
change in sign in the Argonne phase shift just before 50 MeV. }
\label{tab1}
\end{table}

\newpage
%

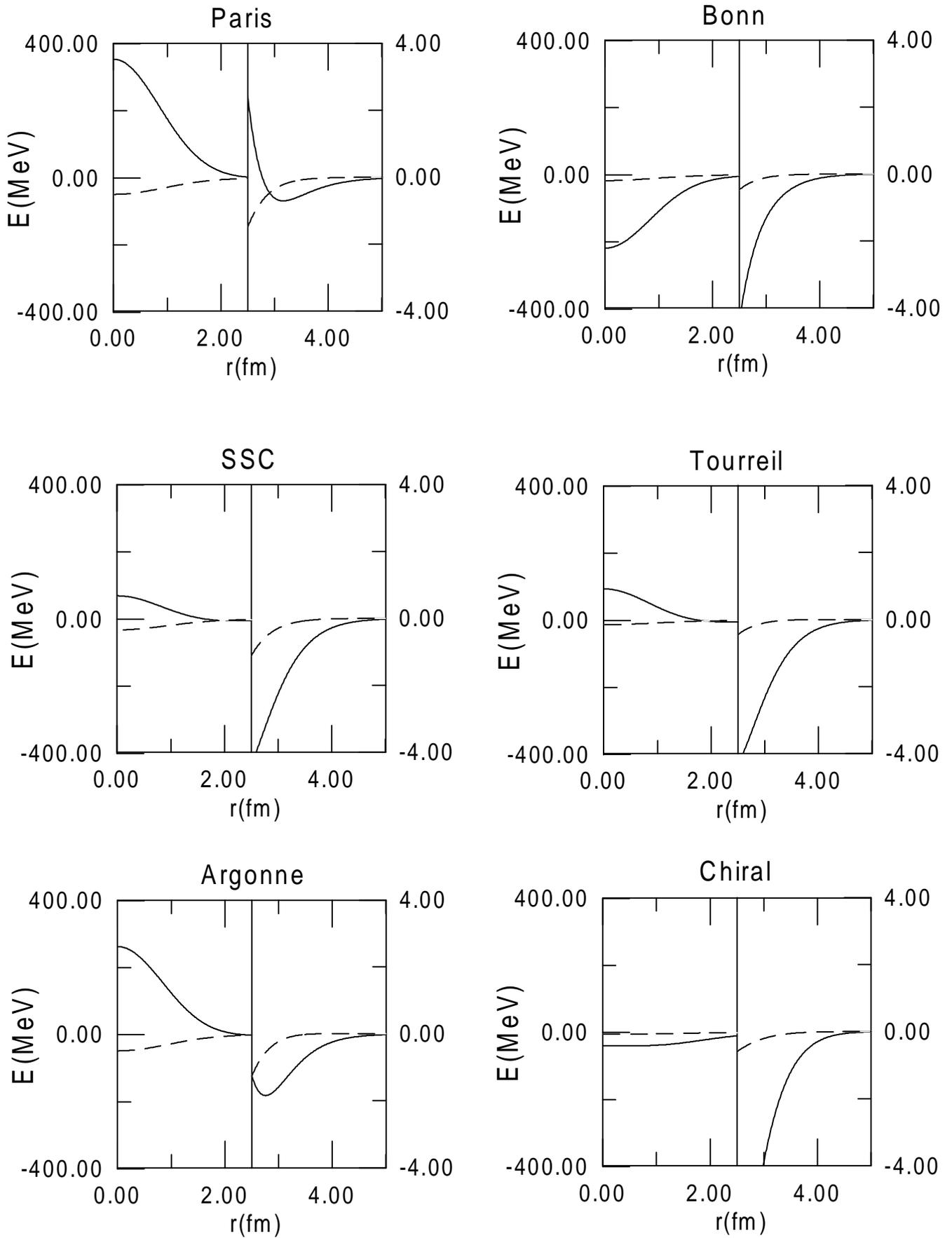
\begin{figure}[tbp]
\caption{Effective $N\alpha $ central (continuous line) and spin-orbit
(dashed line) potentials for the Paris \protect\cite{paris}, Bonn (OBEPR) 
\protect\cite{machleidt1}, supersoft core C\protect\cite{tourreil1}, dTRS 
\protect\cite{tourreil2}, Argonne \protect\cite{wiringa}, and chiral 
\protect\cite{mane2} interactions.}
\label{fig1}
\end{figure}
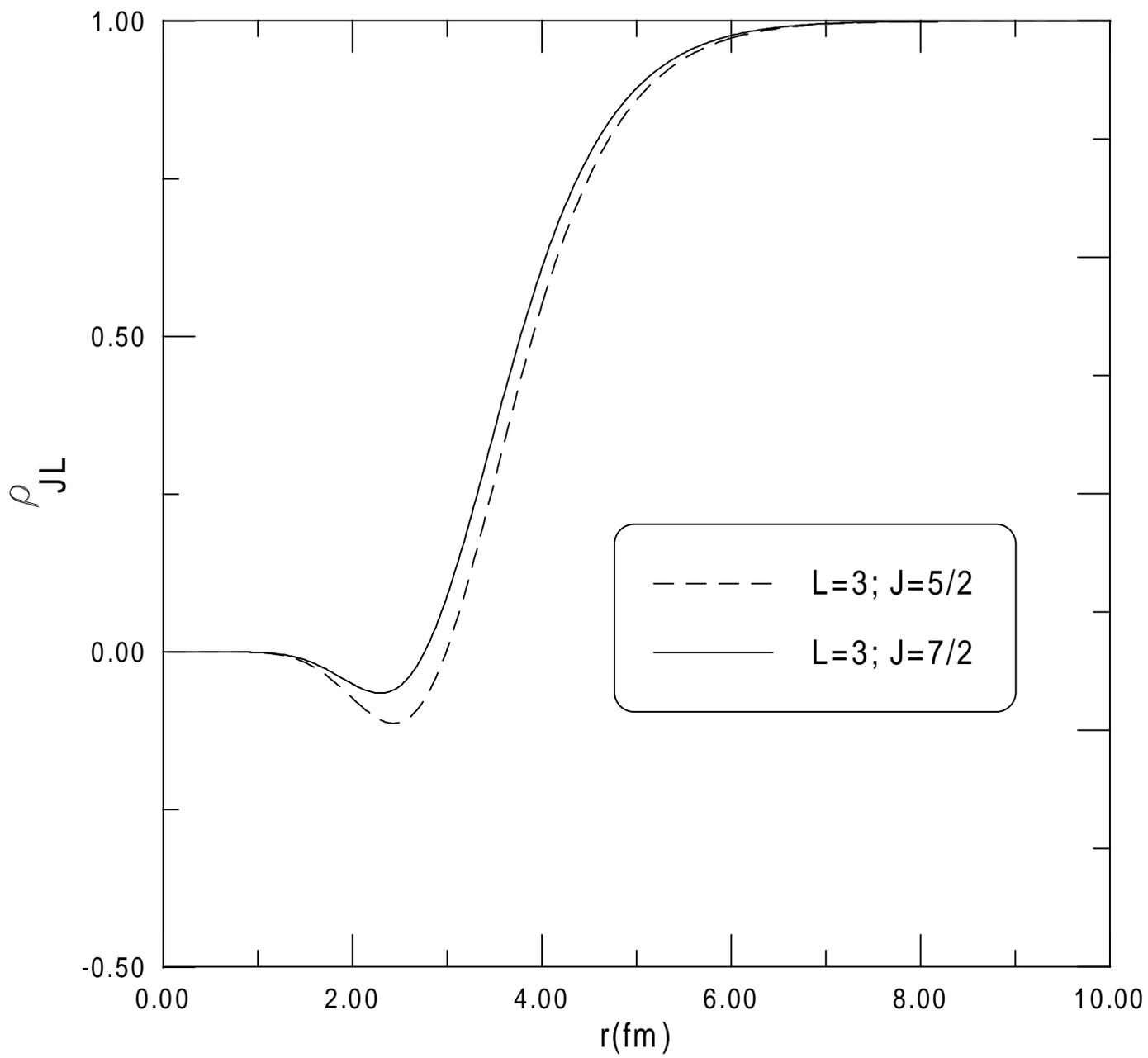
\begin{figure}[tbp]
\caption{Phase shifts normalized to 1 [Eq. \ref{e48}] at 15 MeV, as function
of $r$ for the waves $F_{5/2}$ (dashed line) and $F_{7/2}$ (continuous
line). }
\label{fig2}
\end{figure}
\begin{figure}[tbp]
\caption{$N\alpha $ phase shifts for the Paris (P), Bonn (B), supersoft core
C (S), dTRS (T), Argonne (A), and chiral (C) $NN$ interactions. The
experimental results are taken from Refs. \protect\cite{exper1}($\circ $) , 
\protect\cite{exper4}($\diamond $), \protect\cite{exper3}(+), and 
\protect\cite{exper2}($\times $).}
\label{fig3}
\end{figure}

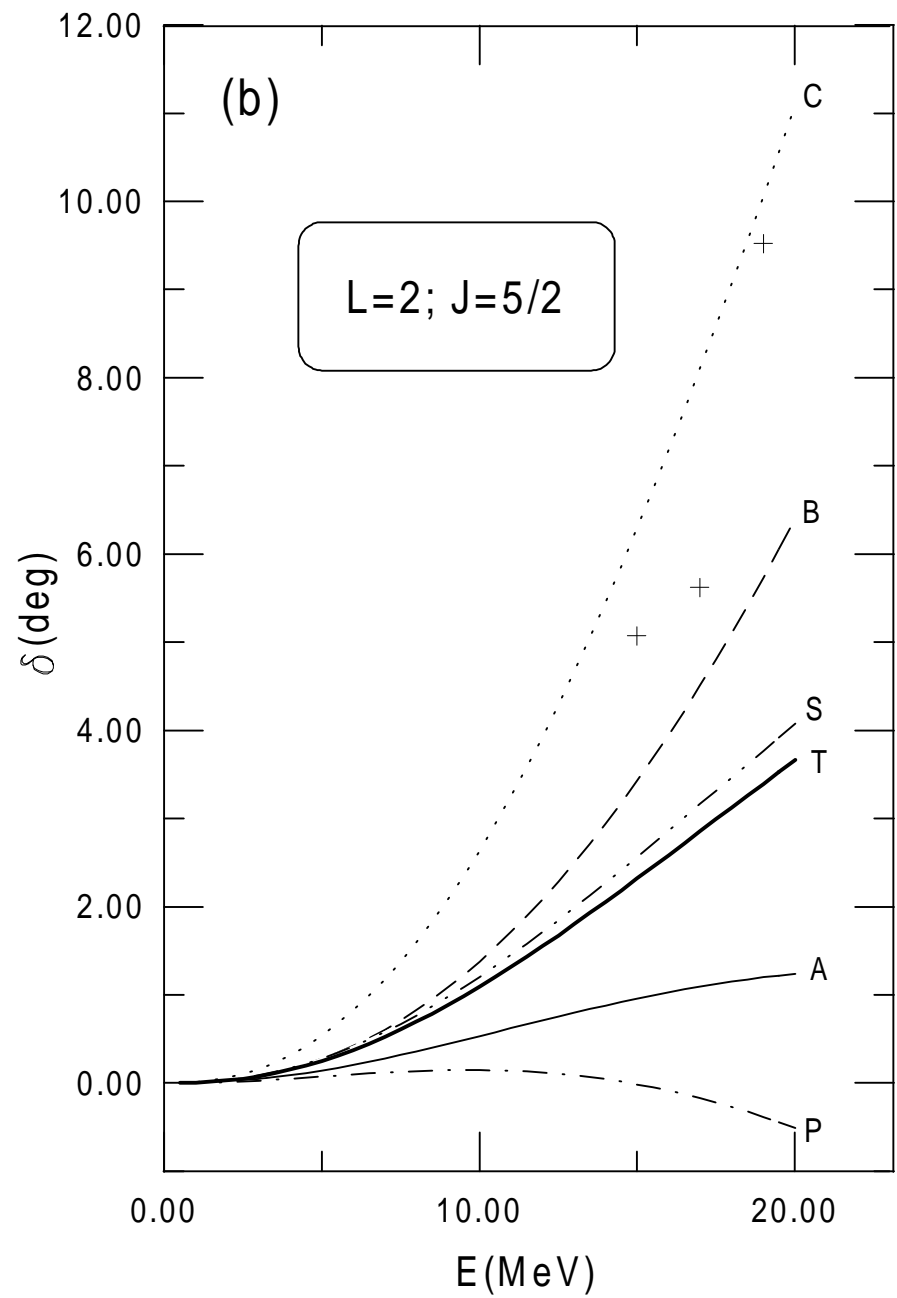
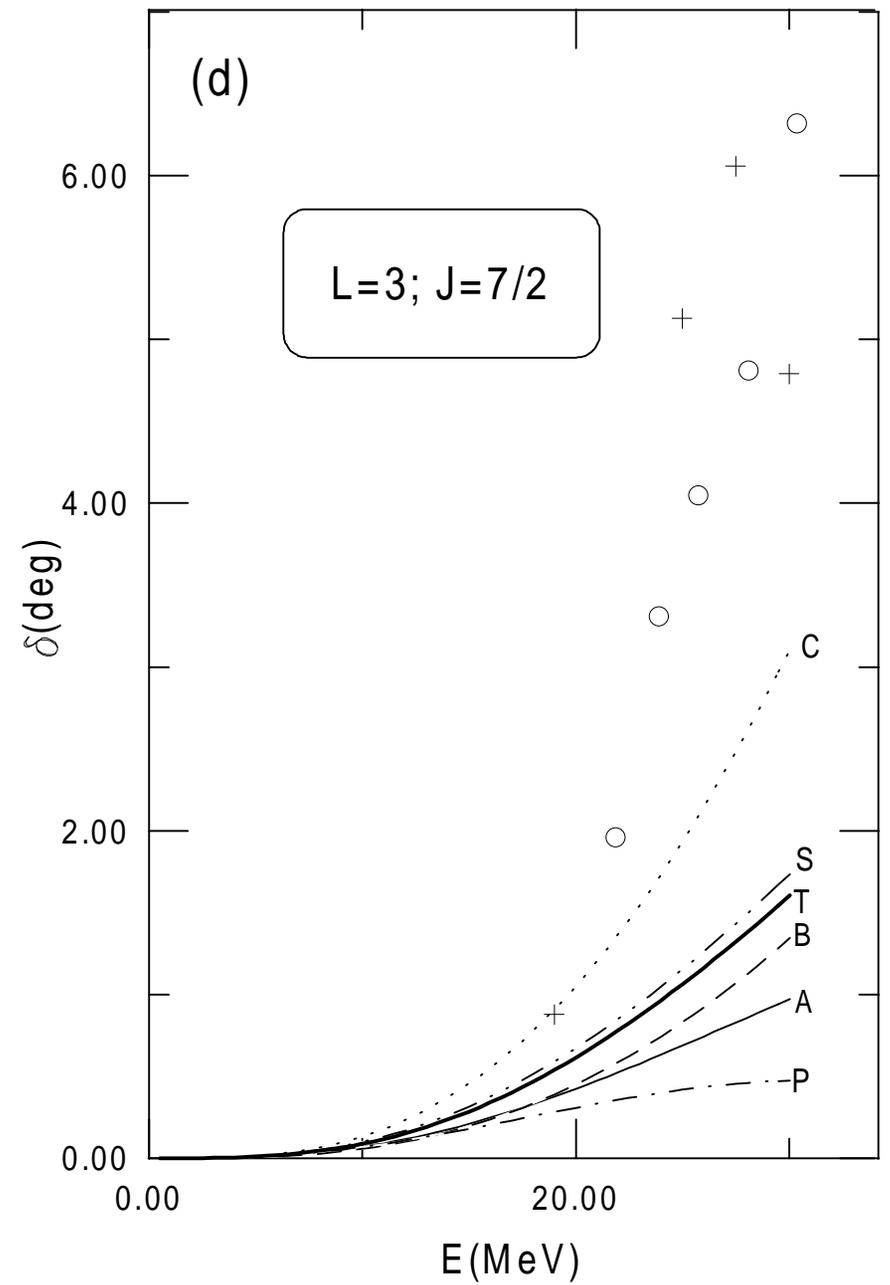
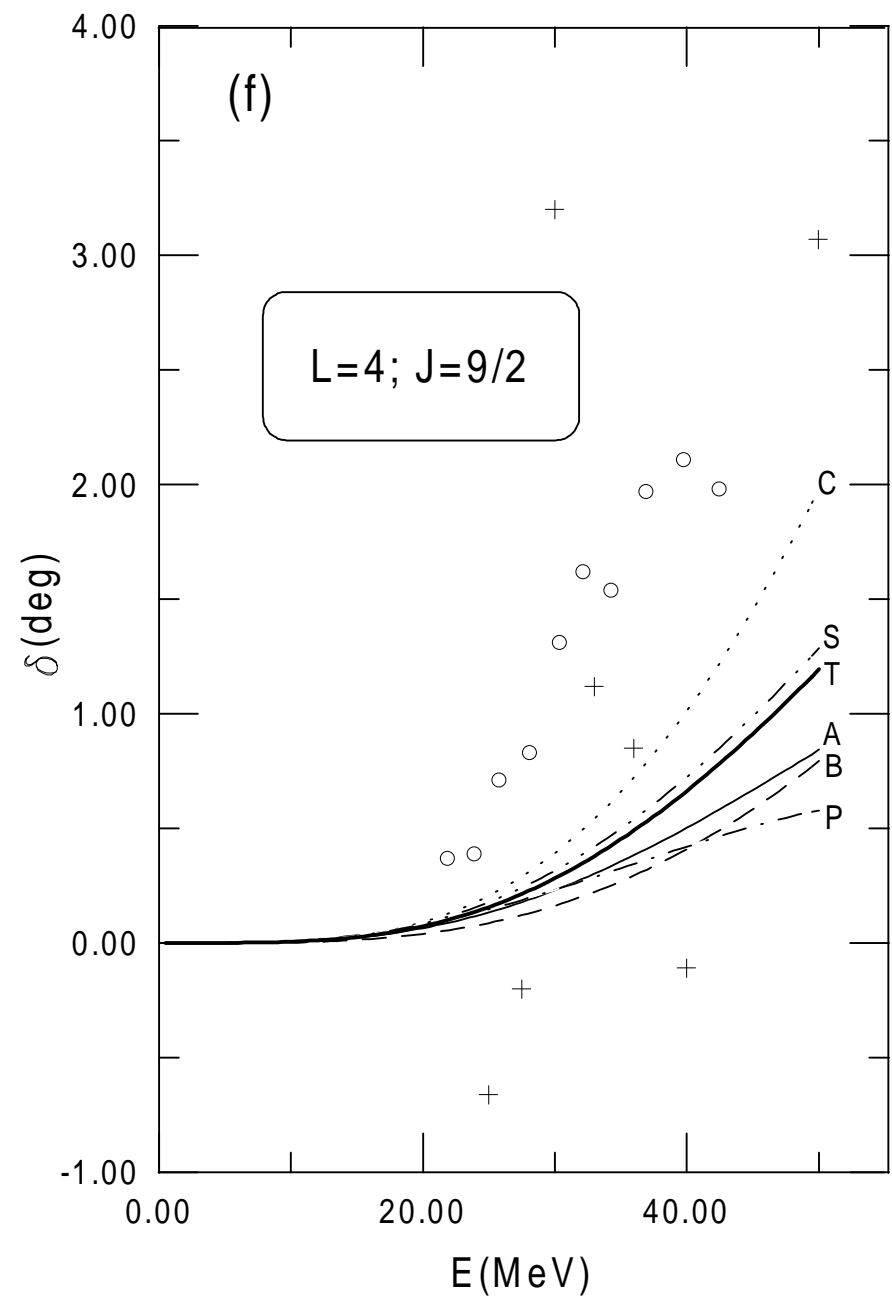
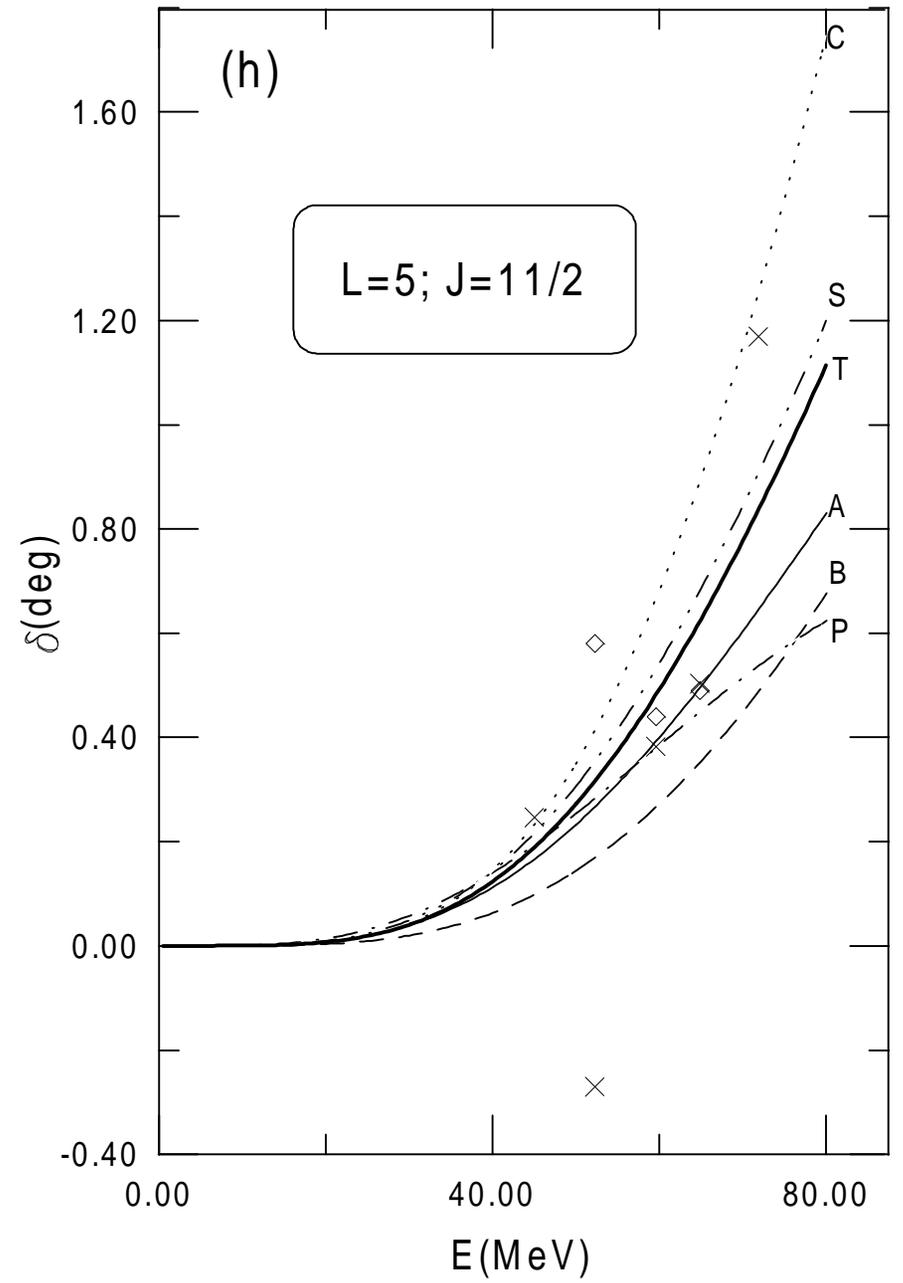
\end{document}